\documentclass[%
 reprint,
 amsmath,amssymb,
 aps,
]{revtex4-1}

\usepackage[sans]{dsfont} 
\usepackage{graphicx}
\usepackage{dcolumn}
\usepackage{bm}
\usepackage{hyperref}

\begin{document}

\preprint{APS/123-QED}

\title{Effective field theory for resonant wino dark matter}
\thanks{This work was presented at the APS DPF 2017 meeting.}%
\author{Evan Johnson}
 \email{johnson.6036@osu.edu}

\author{Hong Zhang}
\email{zhang.5676@osu.edu}

\author{Eric Braaten}
\email{braaten.1@osu.edu}

\affiliation{Department of Physics,
         The Ohio State University, Columbus, OH\ 43210, USA}%


\begin{abstract}
Enhancements in WIMP annihilation rates have been identified with a nonperturbative ``Sommerfeld enhancement'' arising from the exchange of light mediators.
At certain critical values of the WIMP mass, the enhancement is increased dramatically due to a zero-energy resonance at the WIMP-pair threshold.
For masses near these critical values, WIMP scattering processes are accurately described by an effective field theory where WIMPs have nonperturbative zero-range contact interactions.
The effective field theory is controlled by a renormalization-group fixed point at which the WIMPs are degenerate in mass and their scattering length is infinite.
If the WIMPs can exchange massless mediators, the resulting long-range interaction must also be treated nonperturbatively.
We develop an effective field theory for $SU(2)$-triplet dark matter, which have short-range weak and long-range electromagnetic interactions.
We refer to these dark matter particles as {\it winos}.
The long-range Coulomb interaction between charged winos is resummed to all orders. 
The parameters of the effective field theory can be determined by matching wino-wino scattering amplitudes calculated by solving the Schr\"odinger equation with a potential describing exchange of  electroweak gauge bosons.
With Coulomb resummation, the effective field theory at leading order gives a good description of the low-energy two-body observables for winos.
\end{abstract}

\maketitle


\section{Introduction}
\label{sec:Introduction}

Weakly interacting massive particles (WIMPs) are one of the best motivated candidates for a dark matter particle. 
Stable particles with weak interactions with masses roughly at the electroweak scale are naturally produced in the early universe and freeze out with a relic abundance comparable to the observed mass density of dark matter \cite{Kolb:1990vq,Steigman:2012nb}.
If the WIMP mass $M$ is in the TeV range, the self-interactions of nonrelativistic WIMPs are complicated by a nonperturbative effect pointed out by Hisano et al. \cite{Hisano:2002fk}.
Exchanges of gauge bosons must be summed to all orders in the gauge coupling constant and can provide a nonperturbative ``Sommerfeld enhancement'' to low-energy scattering and annihilation cross sections.

For certain critical values of the WIMP mass $M$, the enhancement is increased dramatically due to a zero-energy resonance at the WIMP-pair threshold.
When the resonance is in the S-wave channel, the enhancement in WIMP cross sections occurs over a broader range of $M$ than in higher partial waves.
A feature unique to the S-wave channel is the dynamical generation of a length scale, the scattering length $a$, which can be orders of magnitude larger than the range of the weak interactions.

In a fundamental quantum field theory, WIMPs interact through the exchange of gauge bosons through local interactions.
The Sommerfeld enhancement can be calculated by summing up an infinite number of gauge boson exchange diagrams.
In the Milky Way, WIMPs have velocities on the order of $10^{\text{-}3}$ and can be treated nonrelativistically.
The exchange of weak gauge bosons becomes nonperturbative for relative momentum below $\alpha_2 M$.
Photon exchange becomes nonperturbative for relative momentum below $\alpha M$.
WIMPs with TeV scale masses and relative velocity on the order of $10^{\text{-}3}$ therefore have nonperturbative electroweak interactions.
This challenge is overcome by solving a Schr\"odinger equation in which WIMPs interact via instantaneous interactions at a distance through a potential generated by the exchange of electroweak gauge bosons.
The Sommerfeld enhancement is determined numerically by solving the Schr\"odinger equation with this potential.
We refer to this framework as {\it nonrelativistic effective field theory} (NREFT) and use it as a microscopic description of the WIMP interactions.

Near the critical values of the WIMP mass, the calculation of scattering observables is further facilitated by employing an effective field theory in which WIMPs interact nonperturbativly through contact interactions and charged WIMPs can exchange photons.
We refer to this theory as {\it zero-range effective field theory} (ZREFT).
In Ref.~\cite{Braaten:2017gpq}, we developed the ZREFT for WIMPs that consist of the neutral dark-matter particle $w^0$ and charged WIMPs $w^+$ and $w^-$ with a slightly larger mass.
We refer to these WIMPs as {\it winos}, because the fundamental theory describing them could be the minimal supersymmetric standard model (MSSM) in a region of parameter space where the neutral wino is the lightest supersymmetric particle.

In Ref.~\cite{Braaten:2017gpq}, the ZREFT for winos is developed for winos with short-range weak interactions only.
It is shown to be a systematically improvable effective field theory by calculating wino-wino scattering observables to NLO in the ZREFT power counting.
We build on the results in Ref.~\cite{Braaten:2017gpq} by including the effects of Coulomb resummation in the interactions between charged winos and show that the theory gives good agreement even at leading order, where there is a single free parameter.
This paper serves to summarize the main results of Ref.~\cite{Braaten:2017kci}, presented here without derivation.
In Ref.~\cite{BJZ-Annihilation}, we complete the effective field theory by including the effects of wino-pair annihilation into electroweak gauge bosons.

This paper is organized as follows. 
In Sec.~\ref{sec:QFT}, we present the fundamental relativistic quantum field theory describing winos and their electroweak interactions.
In Sec.~\ref{sec:NREFT}, we describe the nonrelativistic treatment of wino scattering processes by solving a Schr\"odinger equation.
In Sec.~\ref{sec:ZRMC}, we introduce the zero-range model that describes winos with nonperturbative local interactions, and carry out the Coulomb resummation.
In Sec.~\ref{sec:ZREFTC}, we define the zero-range effective field theory by identifying the appropriate renormalization group fixed point and a power counting for perturbations around that fixed point.
In Sec.~\ref{sec:LOpredictions}, we compare results calculated numerically by solving the Schr\"odigner equation with results calculated analytically in ZREFT at leading order.
We conclude in Sec.~\ref{sec:summary}.

\section{Fundamental theory}
\label{sec:QFT}

We assume the dark-matter particle is the neutral member of an $SU(2)$ triplet of Majorana fermions with zero hypercharge. 
The Lorentz-invariant quantum field theory that provides a fundamental  description of these fermions could simply be an extension of the Standard Model with this additional $SU(2)$ multiplet and with a symmetry that forbids the decay of the fermion into Standard Model particles. 
The fundamental theory could also be the Minimal Supersymmetric Standard Model (MSSM) in a region of parameter space where the lightest supersymmetric particle is a wino-like neutralino.
In either case, we refer to the particles in the $SU(2)$ multiplet as {\it winos}. 
We denote the neutral wino by $w^0$ and the charged winos by $w^+$ and $w^-$. 
For the neutral wino to be a stable dark matter candidate, the masses of the neutral and charged winos must be split by a small amount $\delta$. 
In the MSSM, the mass splitting arises from radiative corrections and has a value $\delta=170$~MeV.

The winos can be represented by a triplet of Majorana spinor fields.
We will take the neutral wino mass $M$ to be an adustable parameter, and we keep the wino mass splitting fixed at $\delta= 170$~MeV.
The most important interactions of the winos are those with the electroweak gauge bosons: the photon, the $W^\pm$, and the $Z^0$.
The relevant Standard Model parameters are the mass $m_W = 80.4$~GeV of the $W^\pm$, the mass $m_Z = 91.2$~GeV of the $Z^0$, the $SU(2)$ coupling constant $\alpha_2= 1/29.5$, the electromagnetic coupling constant $\alpha = 1/137.04$, and the weak mixing angle, which is given by $\sin^2 \theta_W = 0.231$.

\section{Nonrelativistic  effective field theory}
\label{sec:NREFT}

Low-energy winos can be described by a nonrelativistic effective field theory in which they interact through potentials that arise from the exchange of weak gauge bosons and in which charged winos also have electromagnetic interactions. 
We call this effective field theory {\it NREFT}. 
In NREFT, the nonrelativistic wino fields are 2-component spinor fields: $\zeta$ which annihilates a neutral wino $w^0$, $\eta$ which annihilates a charged wino $w^-$, and $\xi$ which creates a charged wino $w^+$. 
The kinetic terms for winos in the Lagrangian density are
\begin{eqnarray}
\mathcal{L}_{\rm kinetic} &=& \zeta^\dagger\left(i\partial_0+\frac{\bm{\nabla}^2}{2M}\right) \zeta 
+ \eta^\dagger\left(iD_0+\frac{\bm{D}^2}{2M} -\delta\right) \eta
\nonumber\\
&& + \xi^\dagger\left(iD_0-\frac{\bm{D}^2}{2M} +\delta\right) \xi \;,
\label{eq:kineticLHMN}
\end{eqnarray}
where $D_0$ and $\bm{D}$ are electromagnetic covariant derivatives acting on the charged wino fields.

In refs.~\cite{Hisano:2003ec,Hisano:2004ds}, Hisano, Matsumoto, and Nojiri calculated the nonperturbative effect of the exchange of electroweak gauge bosons between winos on the annihilation rate of a pair of winos into electroweak gauge bosons by solving a Schr\"odinger equation that can be derived from NREFT.
We refer to the channel containing $w^0 w^0$ as the {\it neutral channel} and the channel containing $w^+ w^-$ as the {\it charged channel}. 
We use the indices 0 and 1 to label these channels respectively. The radial Schr\"odinger equation for the S-wave states is
\begin{equation}
\left[ -\frac{1}{M} \mathds{1} \left( \frac{d\ }{dr} \right)^2
+\bm{V}(r) \right] r \binom{R_0(r)}{R_1(r)} = E\,  r \binom{R_0(r)}{R_1(r)} \;,
\label{eq:radialSchrEq}
\end{equation}
where the potential matrix $\bm{V}(r)$ is given as
\begin{equation}
\bm{V}(r) \;=\; \begin{pmatrix} 0 &\; & -\sqrt{2}\alpha_2 \,\frac{e^{-m_W r}}{r} \\ 
-\sqrt{2}\alpha_2 \,\frac{e^{-m_W r}}{r} &\;& 2\delta- 
\frac{\alpha}{r} - \alpha_2 c^2_W\, \frac{e^{-m_Z r}}{r} \end{pmatrix} \;.
\label{eq:V-matrix}
\end{equation}

The scattering thresholds for the neutral and charged channels are $E=0$ and $E= 2\delta$, respectively.
The splitting introduces an important momentum scale associated with transitions between the neutral and charged channels: $\Delta = \sqrt{2M\delta}$.
Solving Eq.~\eqref{eq:radialSchrEq} for the asymptotic wavefunctions determines a dimensionless, unitary, and symmetric S-matrix, $\bm{S}(E) = \mathds{1} + i \, \bm{T}(E)$, where $\bm{T}(E)$ is the dimensionless T-matrix.
At energies $E>2\delta$, both channels are open and the S- and T-matricies are $2 \times 2$. Below the charged channel threshold, only the neutral channel is open.
The S-matrix can be expressed as a phase: $S_{00}(E) = e^{2i\delta_0(E)}$, where $\delta_0(E)$ is the real-valued S-wave phase shift. The T-matrix element $T_{00}(E)$ can be expressed as
\begin{equation}
\frac{2p}{T_{00}} = p \cot\delta_0 -ip \;,
\label{eq:T00pcotdelta}
\end{equation}
where $p=\sqrt{ME}$ is the relative momentum.

Wino-wino elastic cross sections, averaged over initial spins and summed over final spins, can be calculated from the T-matrix as
\begin{subequations}
\begin{eqnarray}
\sigma_{0 \to j}(E) &=& \frac{2\pi}{M^2 v_0(E)^2} \big| T_{j0}(E) \big|^2 \;,
\label{eq:sig0j-T}
\\
\sigma_{1\to j}(E) &=&   \frac{\pi}{M^2 v_1(E)^2}\big| T_{j1}(E) \big|^2 \;,
\label{eq:sig1j-T}
\end{eqnarray}
\label{eq:sigij-T}%
\end{subequations}
where
\begin{subequations}
\begin{eqnarray}
v_0(E) &=&  \sqrt{E/M},
\label{eq:v0-E}
 \\
v_1(E) &=&  \sqrt{(E-2 \delta)/M}.
\label{eq:v1-E}
\end{eqnarray}%
\label{eq:v0,1-E}%
\end{subequations}%
are the wino velocities in the center-of-mass frame with total energy $E$.
The neutral-wino elastic cross section $\sigma_{0 \to 0}$ has a threshold energy of $E=0$.
The charged-wino elastic cross section $\sigma_{1 \to 1}$ and the transition cross sections $\sigma_{0 \to 1}$ and $\sigma_{1 \to 0}$ have thresholds of $2\delta$.

The potential in Eq.~\eqref{eq:V-matrix} can support S-wave resonances at energies dependent on the wino mass $M$.
At certain values of the wino mass, there can be a zero-energy resonance at the neutral-wino-pair threshold.
Near these resonances, $\sigma_{0 \to 0}$ is dramatically enhanced by orders of magnitude.
We refer to a value of the wino mass where the cross section is resonantly enhanced as a {\it unitarity mass} and denote it as $M_*$.
When the wino mass is at a unitarity mass, in the limit of $E \to 0$, the cross section saturates the S-wave unitarity bound for identical spin-$\tfrac12$ particles and we refer to a system where $M=M_*$ as being {\it at unitarity}.
The unitarity bounds for neutral and charged wino elastic scattering are
\begin{subequations}
\begin{eqnarray}
\sigma_{0\to 0}(E)  &\le& \frac{8 \pi}{ME} \;,
\label{eq:sigma-unitarity0}
\\
\sigma_{1\to 1}(E)  &\le& \frac{4 \pi}{M(E-2\delta)} \;.
\label{eq:sigma-unitarity1}
\end{eqnarray}
\label{eq:sigma-unitarity}
\end{subequations}
The factor of two difference between the bounds is because the charged winos are distinguishable.

\begin{figure}[t]
\centering
\includegraphics[width=0.9\linewidth]{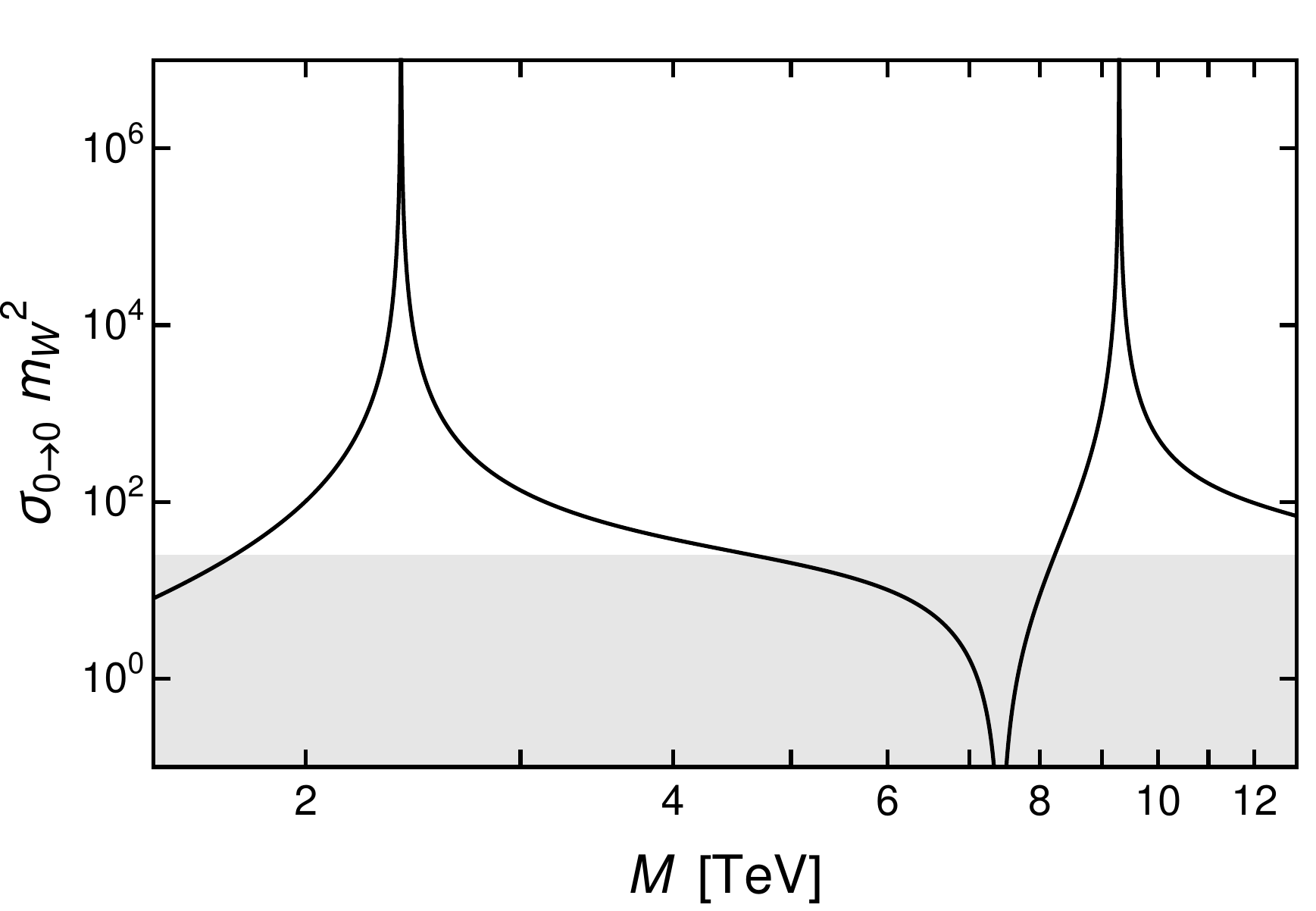}
\caption{Neutral-wino elastic cross section $\sigma_{0 \to 0}$ at zero energy as a function of the wino mass $M$.
The divergent peaks occur at unitarity masses $M_*$ where there is a zero-energy resonance at the neutral-wino-pair threshold and the cross section saturates the unitarity bound.
The first unitarity mass is at 2.39~TeV. 
The range of $M$ where $\sigma_{0 \to 0}$ is above the shaded region ($\sigma_{0 \to 0} < 8 \pi/m_W^2$), is the range of applicability for the zero-range effective field theory, as discussed in Section.~\ref{sec:ZRMC}.
}
\label{fig:sigma00vsM}
\end{figure}

With the mass splitting $\delta$ fixed at 170 MeV, the first unitarity mass is at 2.39 TeV.
The resonant enhancement of the neutral-wino elastic cross section is illustrated in Fig.~\ref{fig:sigma00vsM}, where the cross section is shown as a function of the wino mass.

\begin{figure}[t]
\centering
\includegraphics[width=0.9\linewidth]{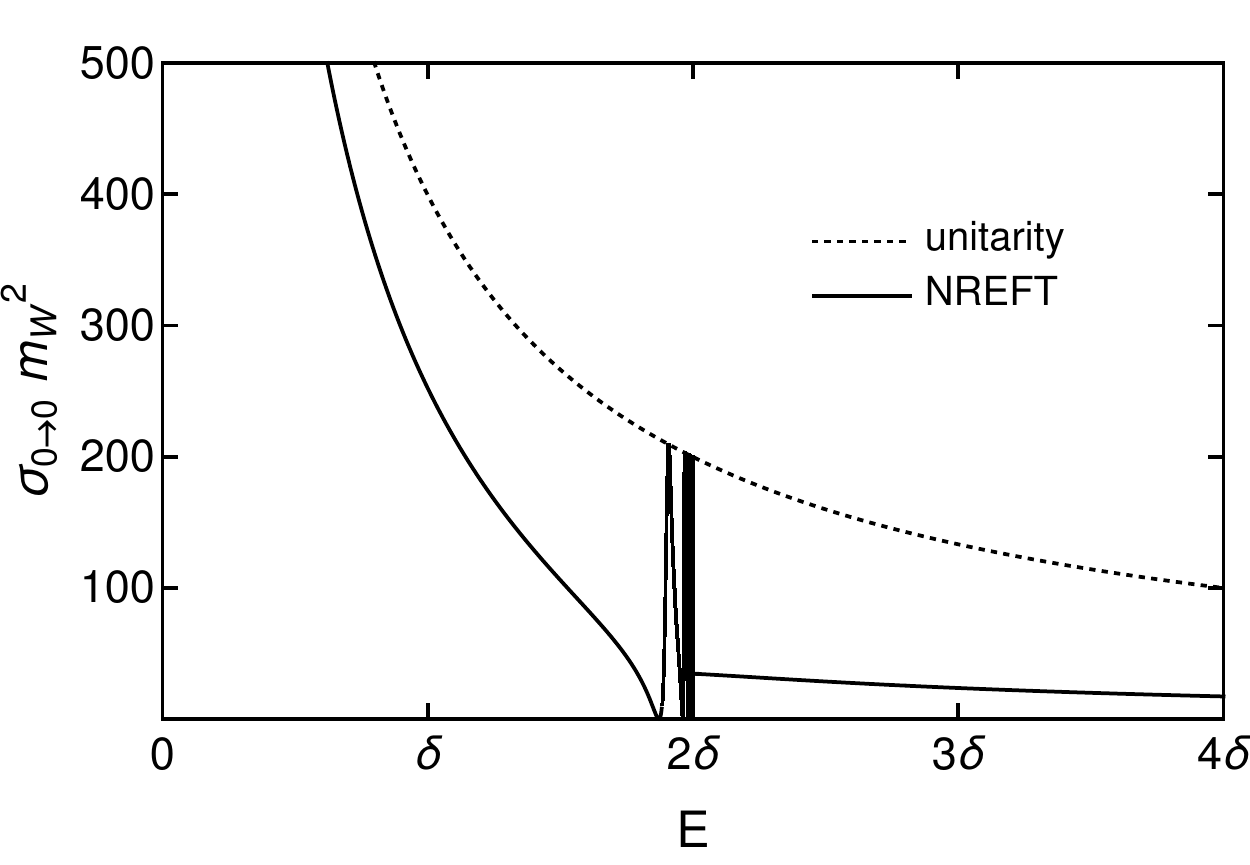}
\caption{Neutral-wino elastic cross section $\sigma_{0 \to 0}$  as a function of the energy $E$ is shown as a solid curve. 
The S-wave unitarity bound is shown as a dotted curve.}
\label{fig:sigma00-NREFT}
\end{figure}

At unitarity, the neutral-wino elastic cross section has dramatic energy dependence. At small energies, the unitarity bound in Eq.~\eqref{eq:sigma-unitarity0} is saturated, and diverges in the $E \to 0$ limit. The cross section and unitarity bound are shown together in Fig.~\ref{fig:sigma00-NREFT}.

Just below the charged-wino-pair threshold at $E= 2\delta$, the cross section has a sequence of narrow resonances whose peaks saturate the unitarity bound. The resonances can be interpreted as bound states in the Coulomb potential for the charged-wino pair $w^+w^-$. Above the threshold at $E= 2\delta$, the cross section is well behaved and slowly decreases as the energy increases.

\begin{figure}[t]
\centering
\includegraphics[width=0.9\linewidth]{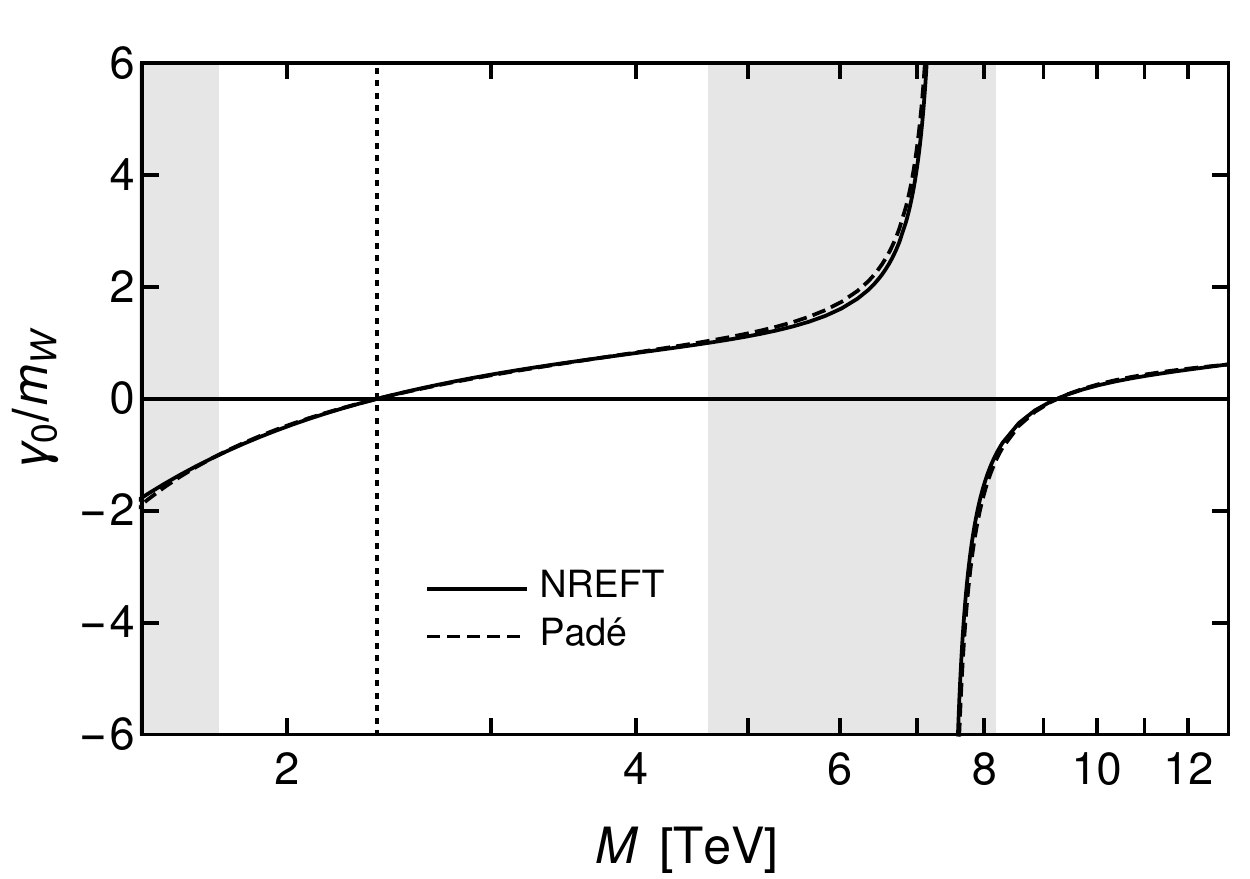}
\caption{The inverse neutral-wino scattering length $\gamma_0$ as a function of the wino mass $M$ (solid curve). 
The dashed curve is the Pad\'e approximant in Eq.~\eqref{eq:gamma0Pade}.
The vertical dotted line indicates the unitarity mass at $M_* = 2.39$~TeV. 
The shaded regions are the ranges of $M$ in which $|\gamma_0| > m_W$, so a zero-range effective field theory is not applicable, as discussed in Section.~\ref{sec:ZRMC}.
}
\label{fig:gamma0_vsM}
\end{figure}

\begin{figure}[t]
\centering
\includegraphics[width=0.9\linewidth]{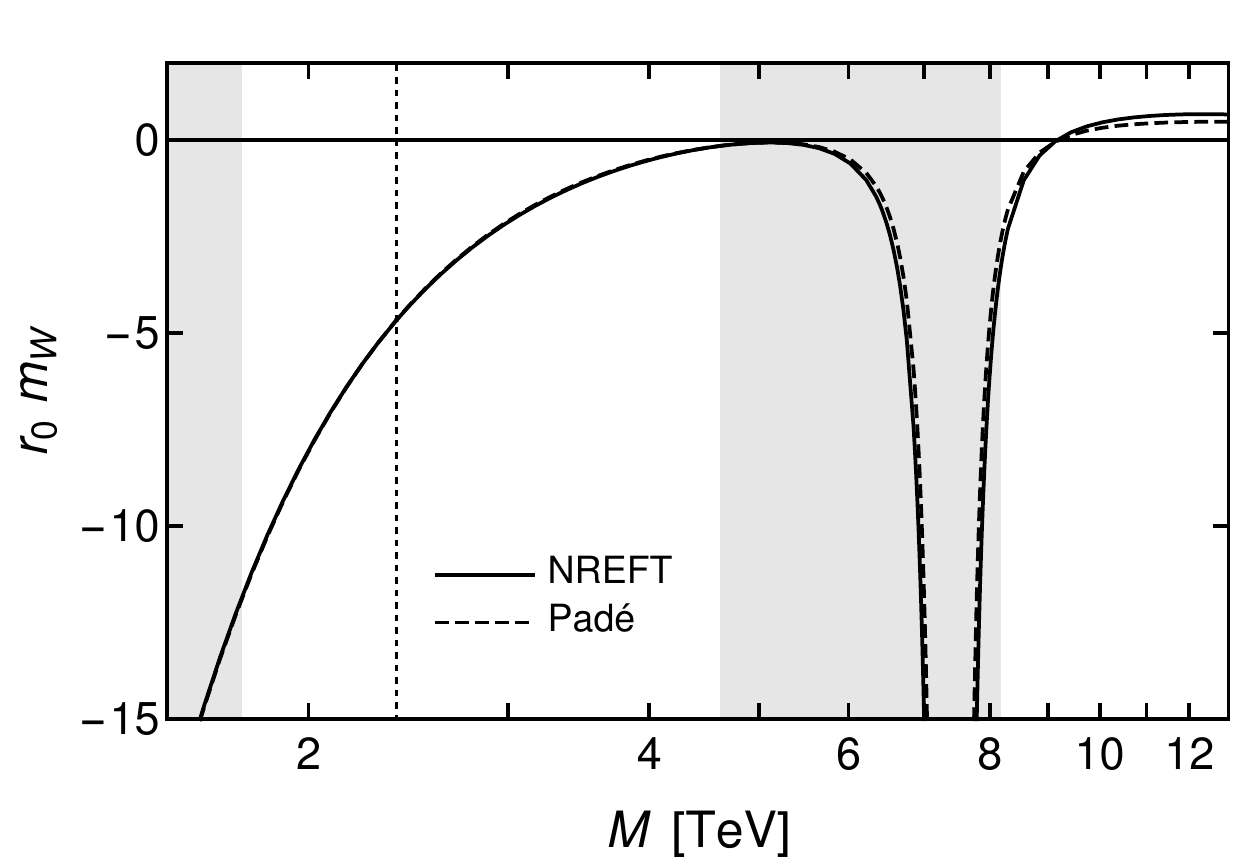}
\caption{The neutral-wino effective range $r_0$ as a function of the wino mass $M$ (solid curve). 
The dashed curve is the Pad\'e approximants in Eq.~\eqref{eq:r0Pade}.
The vertical dotted line indicates the unitarity mass $M_* = 2.39$~TeV.
The shaded regions are the ranges of $M$ in which $|\gamma_0| > m_W$, so a zero-range effective field theory is not applicable, as discussed in Section.~\ref{sec:ZRMC}..
}
\label{fig:r0vsM}
\end{figure}

Neutral winos with energies well below the charged-wino-pair threshold have short-range interactions, because the Coulomb interaction enters only through virtual charged winos. The short-range interactions guarantee that the expression $2p/T_{00}$ in the left hand side of Eq.~\eqref{eq:T00pcotdelta} can be expanded in powers of the relative momentum $p = \sqrt{ME}$:
\begin{equation}
\frac{2p}{T_{00}} = -\gamma_0 -ip + \frac12 r_0\,p^2 + \frac18 s_0\,p^4 + {\cal O}(p^6) \;.
\label{eq:T00NRinv}
\end{equation}
This defines the {\it effective range expansion}.
The $-ip$ term on the right hand side is the only imaginary part, and only odd power of the relative momentum $p$.
The coefficient $-i$ is determined by unitarity of the S-matrix.
The coefficients of ${(p^2)}^n$ are real valued and define observable scattering parameters: $\gamma_0$ is the {\it inverse neutral-wino S-wave scattering length}, $r_0$ is the {\it effective range}, and $s_0$ is the {\it shape parameter}.
The coefficients in the effective range expansion can be determined numerically by solving the Schr\"odinger equation.
The inverse scattering length $\gamma_0$ and the effective range $r_0$ are shown as a function of the wino mass $M$ in Fig.~\ref{fig:gamma0_vsM} and Fig.~\ref{fig:r0vsM}.

The mass dependence on the inverse scatting length $\gamma_0$ and the effective range $r_0$ can be fit accurately by Pad\'e approximants.
The inverse scattering length can be fit by a [2,2] Pad\'e.
The effective range has an offset equal to the local maximum near 5 TeV.
Once this offset is subtracted, the remainder can be fit with a [3,4] Pad\'e.
The Pad\'e approximations are
\begin{subequations}
\begin{eqnarray}
\gamma_0(M) &=& (1.05\, m_W) \, \frac{(M-M_*)(M-M_*')}{(M-M_0)(M-M_0')} \;,
\label{eq:gamma0Pade}
\\
r_0(M) &=& \frac{4.88}{m_W} \left( \frac{M_*(M-M')^2 (M-M'')}{(M-M_0)^2(M-M_0')^2} - 0.0113 \right)
\nonumber\\
\label{eq:r0Pade}
\end{eqnarray}%
\label{eq:gamma0r0Pades}%
\end{subequations}%
The zeros and poles in Eq.~\eqref{eq:gamma0Pade} are $M_*=2.39$~TeV, $M_*'=9.23$~TeV, $M_0=0.845$~TeV, and $M_0'=7.39$~TeV. 
In Eq.~\eqref{eq:r0Pade}, the zeros and poles are $M'=5.13$~TeV, $M''=9.11$~TeV, $M_0=0.129$~TeV, and $M_0'=7.39$~TeV. 
These results will be used in the Section~\ref{sec:ZREFTC} to fix the parameters of the zero-range effective field theory.

\section{Zero-Range Model with Coulomb Resummation}
\label{sec:ZRMC}

The important momentum scales for wino interactions discussed thus far are momentum scales associated with weak exchange: $\alpha_2M$, photon exchange: $\alpha M$, and transitions between channels: $\Delta = \sqrt{2M\delta}$.
At the unitarity mass $M_*=2.39$~TeV, these scales are 81.1~GeV, 17.5~GeV, and 28.5~GeV, respectively.
The S-wave resonance at the neutral-wino-pair threshold generates another small momentum scale, the inverse S-wave scattering length $\gamma_0$ \cite{Braaten:2004rn}.
The mass dependence of $\gamma_0$ is shown in Fig.~\ref{fig:gamma0_vsM}.
It vanishes at the unitarity mass and its absolute value remains smaller than $m_W$, the inverse range of weak interactions, for wino mass in the range $1.75-4.6$~TeV.
In this range, wino interactions can be described by zero-range contact interactions and long-range electromagnetic interctions.
The contact and long-range interactions must be treated nonperturbativly in order to generate the small momentum scale $\gamma_0$.

A simple nonrelativistic field theory for low-energy winos with local interactions is the {\it zero-range model} introduced in Ref.~\cite{Braaten:2017gpq}.
The winos are described by nonrelativistic two-component spinor fields $w_0$, $w_+$, and $w_-$ that annihilate $w^0$, $w^+$, and $w^-$, respectively.
They can be identified with the fields $\zeta$, $\xi^\dagger$, and $\eta$ in NREFT, respectively.
The kinetic terms for winos in the Lagrangian for the zero-range model are
\begin{eqnarray}
\mathcal{L}_{\rm kinetic} &=& w_0^\dagger\left(i\partial_0+\frac{\bm{\nabla}^2}{2M}\right) w_0  
\nonumber\\
&& + \sum_\pm w_\pm^\dagger\left(iD_0+\frac{\bm{D}^2}{2M}-\delta\right) w_\pm \;.  
\label{eq:kineticL}
\end{eqnarray}
The electromagnetic covariant derivatives are
\begin{eqnarray}
D_0 w_\pm &=& (\partial_0 \pm ieA_0)w_\pm \;,
\nonumber\\
\bm{D} w_\pm &=& (\bm{\nabla} \mp ie\bm{A})w_\pm \;.
\end{eqnarray}
Neutral and charged winos have the same kinetic mass $M$, with the mass splitting $\delta$ taken into account through the rest energy of the charged winos.
Since neutral winos are Majorana fermions, they can only have S-wave interactions in the spin-singlet channel.
That channel is coupled to the spin-singlet channel for charged winos.
The Lagrangian for zero-range interactions in the spin-singlet channel can be expressed as
\begin{eqnarray}
\mathcal{L}_{\rm zero-range} &=& 
-\tfrac{1}{4} \lambda_{00} ( w_0^{c\dagger} w_0^{d\dagger} )
\tfrac12 ( \delta^{ac}\delta^{bd}- \delta^{ad}\delta^{bc}) ( w_0^a w_0^b )
\nonumber\\
&& 
-\tfrac{1}{2} \lambda_{01} (  w_+^{c\dagger} w_-^{d\dagger} )
\tfrac12 ( \delta^{ac}\delta^{bd}- \delta^{ad}\delta^{bc}) ( w_0^a w_0^b )
\nonumber\\
&& 
-\tfrac{1}{2} \lambda_{01} (  w_0^{c\dagger} w_0^{d\dagger} )
\tfrac12 ( \delta^{ac}\delta^{bd}- \delta^{ad}\delta^{bc}) ( w_+^a w_-^b )
\nonumber\\
&& 
- \lambda_{11} ( w_+^{c\dagger} w_-^{d\dagger} )
\tfrac12 ( \delta^{ac}\delta^{bd}- \delta^{ad}\delta^{bc})
( w_+^a w_-^b ) \;,
\nonumber\\
\label{eq:ZRint}
\end{eqnarray}
where $\lambda_{00}$, $\lambda_{01}$, and $\lambda_{11}$ are real-valued bare coupling constants.
The factor $\frac12( \delta^{ac}\delta^{bd}- \delta^{ad}\delta^{bc})$ is the projector onto the spin-singlet channel.

In the zero-range model, the S-wave spin-singlet transition amplitudes $\mathcal{A}_{ij}(E)$ are functions of the total energy $E$ of the wino pair only.
The T-matrix elements $\mathcal{T}_{ij}(E)$ for  S-wave wino-wino scattering are obtained by evaluating the transition amplitudes $\mathcal{A}_{ij}(E)$ on the energy shell.  
The constraints on the T-matrix elements from S-wave unitarity can be derived from the unitarity condition for the amplitude matrix $\bm{\mathcal{A}}(E)$ at real $E$, which can be expressed as
\begin{eqnarray}
\bm{\mathcal{A}}(E)- \bm{\mathcal{A}}(E)^* &=& 
- \frac{1}{8 \pi} \bm{\mathcal{A}}(E) \bm{M}^{1/2} \Big[ \bm{\kappa}(E)  - \bm{\kappa}(E)^* \Big] 
\nonumber
\\
&&\times \;
\bm{M}^{1/2} \bm{\mathcal{A}}(E)^* \;,
\label{eq:A-unitarity}
\end{eqnarray}
where $\bm{M}$ is the $2\times2$ diagonal matrix
\begin{equation}
\label{eq:Mmatrix}
\bm{M}=
\begin{pmatrix}  M &   0   \\ 
                          0  & 2M   
\end{pmatrix}
\end{equation}
and $\bm{\kappa}$ is a diagonal matrix whose entries are functions of $E$:
\begin{equation}
\bm{\kappa}(E) =
\begin{pmatrix} \kappa_0(E)  &          0       \\ 
                                 0            & \kappa_1(E)
\end{pmatrix} .
\label{eq:kappamatrix}
\end{equation}
The functions $\kappa_0$ and $\kappa_1$ of the complex energy $E$ have branch cuts at 0 and $2\delta$, respectively:
\begin{subequations}
\begin{eqnarray}
\kappa_0(E) &=& \sqrt{-ME-i\varepsilon},
\label{eq:kappa0}
\\
\kappa_1(E) &=& \sqrt{-M(E-2\delta)-i\varepsilon}.
\label{eq:kappa1}
\end{eqnarray}
\label{eq:kappa01}%
\end{subequations}
The different diagonal entries of the matrix $\bm{M}$ in Eq.~\eqref{eq:Mmatrix} are a convenient way to take into account that the neutral channel $w^0w^0$ consists of a pair of identical fermions while the charged channel $w^+w^-$ consists of two distinguishable fermions.

\begin{figure}[t]
\centering
\includegraphics[width=0.9\linewidth]{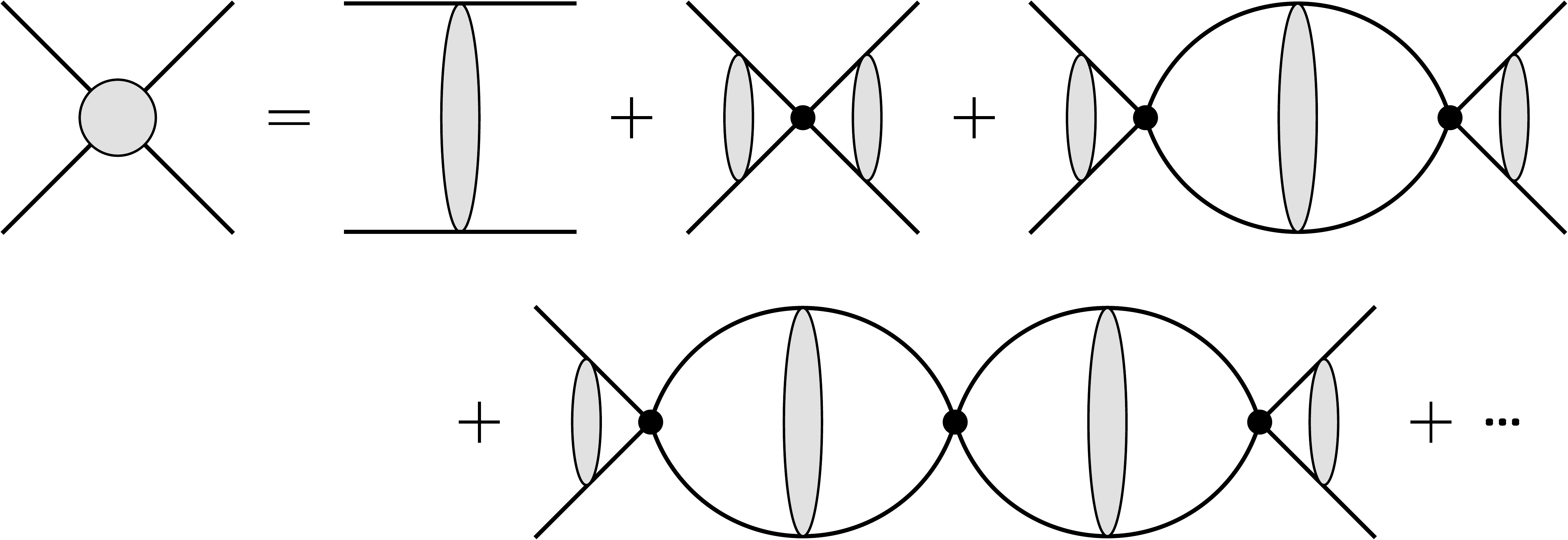}
\caption{Diagrams for the transition amplitudes $\mathcal{A}_{ij}(E)$ expressed as a sum over the number of zero-range interactions. 
A solid line represents either a neutral wino or a charged wino.
A shaded blob on the right side represents Coulomb resummation.
Each bubble is summed over a neutral-wino pair $w^0 w^0$ and a charged-wino pair $w^+ w^-$.
The bubble diagrams must be summed to all orders.
}
\label{fig:ZREFTSumC}
\end{figure}

The amplitude ${\cal A}_{ij}(E)$ with Coulomb resummation is given by the sum of all diagrams with the appropriate incoming and outgoing pairs of wino lines specified by $i$ and $j$ (0 for $w^0w^0$, 1 for $w^+w^-$), and with intermediate zero-range interactions and/or exchanges of Coulomb photons. 
In Fig.~\ref{fig:ZREFTSumC}, the amplitude is expressed as a sum over the number of zero-range interactions.
The pair of incoming wino lines or outgoing wino lines is either $w^0w^0$ or $w^+ w^-$.
Adjacent zero-range vertices are connected by a bubble whose upper and lower wino lines are summed over $w^0 w^0$ and $w^+ w^-$.
The bubble diagrams must be summed to all orders.

\begin{figure}[t]
\centering
\includegraphics[width=0.9\linewidth]{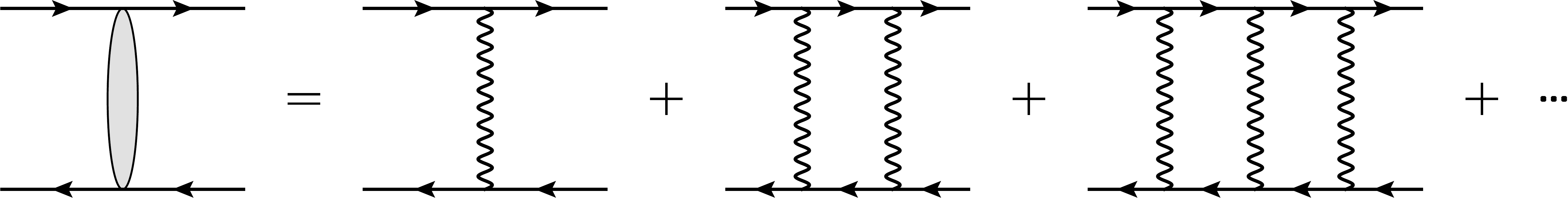}
\\[0.5cm]
\includegraphics[width=0.9\linewidth]{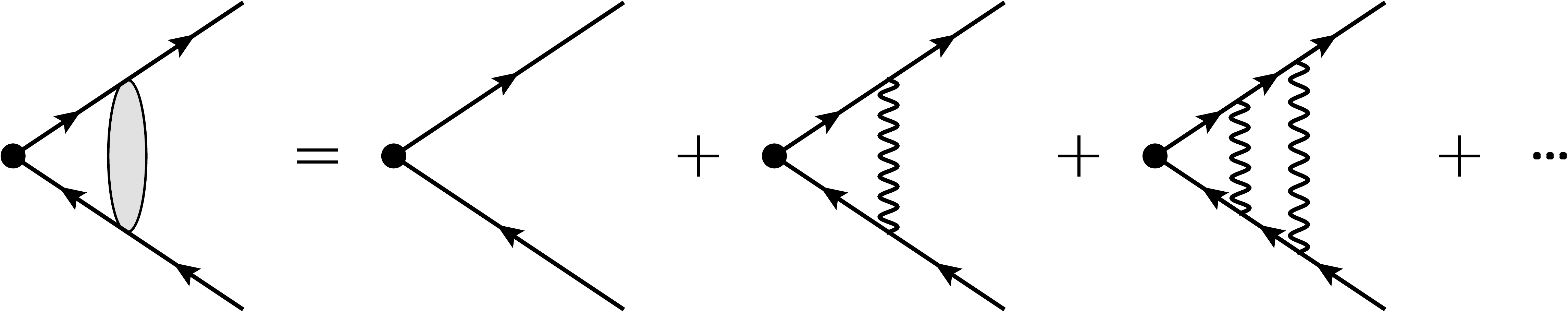}
\\[0.5cm]
\includegraphics[width=0.9\linewidth]{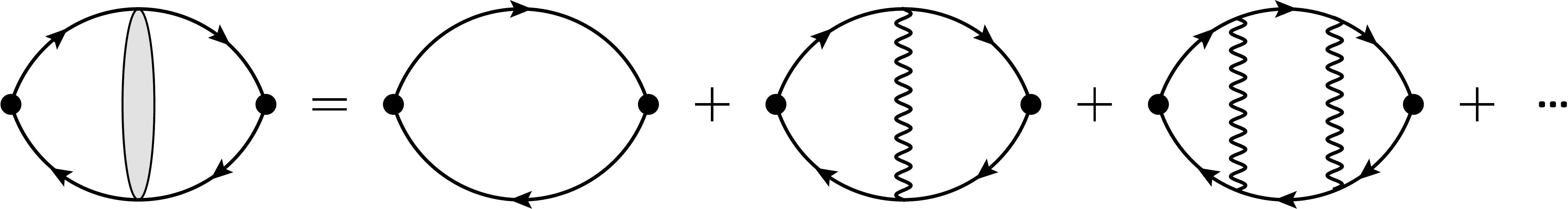}

\caption{Feynman diagrams illustrating Coulomb resummation. 
Top row represents the amplitude ${\cal A}_C$, for $w^+ w^- \to w^+ w^-$ with no zero-range interactions.
The middle row represents the amplitude $W_1$, to create a charged wino pair at a point.
The bottom row represents the virtual charged-wino bubble amplitude.
The exchanges of photons must be summed to all orders.}
\label{fig:CoulombResum}
\end{figure}

Each pair of upper and lower lines in Fig.~\ref{fig:ZREFTSumC} is connected by a blob that represents the sum of all ladder diagrams with the exchange of Coulomb photons.
If the pair of lines is $w^0w^0$, the Coulomb exchange diagrams are 0.
If the pair of lines  in the first diagram on the right side of Fig.~\ref{fig:ZREFTSumC} is $w^+w^-$, the blob represents the sum of Coulomb exchange diagrams in the top line of Fig.~\ref{fig:CoulombResum}.
We have represented the propagators for the charged winos $w^+$ and $w^-$ by solid lines with a forward arrow and a backward arrow, respectively.  
If the pair of outgoing (incoming) lines in any of the remaining diagrams on the right side of Fig.~\ref{fig:ZREFTSumC} is $w^+w^-$, the blob connecting those lines represents the sum of (the complex conjugates of) Coulomb exchange diagrams in the middle row of Fig.~\ref{fig:CoulombResum}.
If the pair of lines  in any bubble is $w^+w^-$, the blob represents the sum of Coulomb exchange diagrams in the bottom row of Fig.~\ref{fig:CoulombResum}.

The matrix of S-wave transition amplitudes for wino pairs with total energy $E$ can be expressed in the form
\begin{equation}
\label{eq:Amatrix}
\bm{\mathcal{A}}(E) =
\bm{\mathcal{A}}_C(E) + \bm{W}(E) \,\bm{\mathcal{A}}_s(E)\, \bm{W}(E) \;.
\end{equation}
The first term on the right side is the matrix amplitude for Coulomb scattering, whose only nonzero entry is the second diagonal entry for $w^+ w^-$ scattering:
\begin{equation}
\label{eq:ACmatrix}
\bm{\mathcal{A}}_C(E) =
\begin{pmatrix} ~0~  & 0  \\  0  & \mathcal{A}_C(E) \end{pmatrix} \;,
\end{equation}
where $\mathcal{A}_C(E)$ is the S-wave Coulomb transition amplitude. 
It is given by the sum of diagrams in the top row of Fig.~\ref{fig:CoulombResum}:
\begin{equation}
\label{eq:ACoulomb}
\mathcal{A}_C(E) = 
\left( 1 - \frac{\Gamma(1+i \eta)}{\Gamma(1-i \eta)}  \right) \frac{2\pi}{M\, \kappa_1(E)} \;,
\end{equation}
where $\kappa_1$ is defined in Eq.~\eqref{eq:kappa1} and $\eta$ is another energy variable defined by
\begin{equation}
\label{eq:eta-def}
\eta(E) \equiv 
 \frac{i\, \alpha \, M}{2\, \kappa_1(E)} 
 = i \frac{\alpha M}{2} \big[ - M(E-2 \delta) - i \epsilon \big]^{-1/2} \;.
\end{equation}
For a real energy $E = 2 \delta + p^2/M$ above the charged-wino-pair threshold, $\eta$ is real and negative: $\eta = - \alpha M/2p$.
For a real energy $E$ below the charged-wino-pair threshold, $\eta$ is pure imaginary.
The amplitude in Eq.~\eqref{eq:ACoulomb} has poles in $E$  at real energies $E_n$ that correspond to Coulomb bound states of $w^+ w^-$: 
\begin{equation}
\label{eq:E-n}
E_n = 2 \delta
- \frac{ \alpha^2 M}{4 n^2} \;,
\end{equation}
where $n$ is a positive integer. 
The matrix $\bm{W}(E)$ in Eq.~\eqref{eq:Amatrix} is diagonal:
\begin{equation}
\label{eq:Wmatrix}
\bm{W}(E)= 
\begin{pmatrix} ~~1~~   & 0 \\  0 & W_1(E) \end{pmatrix} \;.
\end{equation}
Its second diagonal entry $W_1(E)$ is the dimensionless amplitude for $w^+w^-$ created at a point to produce $w^+w^-$ with total energy $E$ in the presence of Coulomb interactions.
It can be obtained diagrammatically by expressing the sum of diagrams in the middle row of Fig.~\ref{fig:CoulombResum} as the leading order vertex multiplied by $W_1(E)$, given by
\begin{equation}
W_1(E) = C(E) \, \left( \frac{\Gamma(1 + i \eta)}{\Gamma(1 - i \eta)} \right)^{1/2} \;,
\label{eq:W1-eta}
\end{equation}
where $\eta$ is the function of $E$ in Eq.~\eqref{eq:eta-def} and  $C$ is the square root of the Sommerfeld factor:
\begin{equation}
C^2(E)  = \frac{2 \pi \eta}{\exp(2 \pi \eta) - 1} \;.
\label{eq:Sommerfeld}
\end{equation}

The matrix  $\bm{\mathcal{A}}_s(E)$ in Eq.~\eqref{eq:Amatrix} is a matrix of short-distance transition amplitudes.
It is the contribution to $\bm{\mathcal{A}}(E)$ from diagrams in which the first interaction and the last interaction are both zero-range interactions.
It can be expressed most simply by giving its inverse:
\begin{equation}
\label{eq:Ainverse0C}
\bm{\mathcal{A}}_s^{-1}(E) = \frac{1}{8\pi} \bm{M}^{1/2}
\Big[ - \bm{\gamma} + \bm{K}(E) \Big] \bm{M}^{1/2} \;,
\end{equation}
where  $\bm{\gamma}$ is a symmetric matrix of renormalized parameters,
\begin{equation}
\label{eq:gammamatrix}
\bm{\gamma}= 
\begin{pmatrix} \gamma_{00}   & \gamma_{01} \\ 
 \gamma_{01} & \gamma_{11} 
\end{pmatrix},
\end{equation}
and $\bm{K}$ is a diagonal matrix that depends on $E$:
\begin{equation}
\bm{K}(E) =
\begin{pmatrix} \kappa_0(E)  &          0       \\ 
                                 0            & K_1(E)
\end{pmatrix} .
\label{eq:Kmatrix}
\end{equation}
Its first diagonal entry is the function $\kappa_0$ in Eq.~\eqref{eq:kappa0}, and its second diagonal entry is
\begin{equation}
K_1(E) =\alpha M \left[ \psi(i \eta) + \frac{1}{2 i \eta} - \log(-i\eta)  \right],
\label{eq:K1-E}
\end{equation}
where $\psi(z)=(d/dz)\log \Gamma(z)$ and $\eta(E)$ is defined in Eq.~\eqref{eq:eta-def}.
This result was first derived by Kong and Ravndal in an application to nuclear physics \cite{Kong:1999sf}.
The matrix $\bm{\mathcal{A}}(E)$ of transition amplitudes in Eq.~\eqref{eq:Amatrix} satisifies the unitarity condition in Eq.~\eqref{eq:A-unitarity} provided the parameters $\gamma_{ij}$ in the matrix $\bm{\gamma}$ in Eq.~\eqref{eq:gammamatrix} are real valued.

\section{ZREFT with Coulomb resummation}
\label{sec:ZREFTC}

The zero-range model with transition amplitudes given by the matrix in Eq.~\eqref{eq:Amatrix} has two coupled scattering channels with different energy thresholds.
An effective field theory can be defined most rigorously by identifying a renormalization-group fixed point. 
Systematically improving accuracy is then ensured by adding to the Lagrangian operators  with increasingly higher scaling dimensions.

More convienently, the matrix $\bm{\gamma}$ in Eq.~\eqref{eq:Ainverse0C} can be expressed as a power series in $p^2$ with coefficients that define scattering parameters.
Systematic improvement can be achieved if we can identify a power counting that determines the improvement in the accuracy of a model that can be obtained by including each of the parameters.  
To define the power counting, we introduce the generic momentum scale $Q$ described by the effective theory. 
We take the energy $E$ and the mass splitting $\delta$ to be of order $Q^2/M$. 
We also introduce a momentum scale $\Lambda$ that can be regarded as the ultraviolet cutoff of the effective field theory. 
The physical interpretation of $\Lambda$ is  the smallest momentum scale beyond the domain of applicability of the effective field theory. 
In the case of winos, $\Lambda$ is the inverse range $m_W$ of the weak interactions. 
The power-counting scheme identifies how each parameter scales as a power of $Q$ and $\Lambda$. 
The transition amplitudes $\mathcal{A}_{ij}(E)$ can be expanded in powers of $Q/\Lambda$. 
The systematically improving accuracy of the effective field theory is ensured by including parameters whose leading contributions to $\mathcal{A}_{ij}(E)$ scale with increasingly higher powers of $Q/\Lambda$.

\subsection{Renormalization-group fixed points}
\label{sec:RG}

Lensky and Birse have carried out a careful renormalization-group (RG) analysis of the two-particle sector for the field theory with two coupled scattering channels with zero-range interactions \cite{Lensky:2011he}. 
Three distinct RG fixed points were identified corresponding to zero, one, or two resonances at the scattering threshold.
For wino scattering, the appropriate fixed point has a single resonance, and we refer to it as the {\it single-channel-unitarity fixed point}. 
At this fixed point, the two scattering channels have the same threshold  ($\delta = 0$), and are mixed by an angle $\phi$.
There is nontrivial scattering in a single channel that is a linear combination of the neutral channel $w^0w^0$ and the charged channel $w^+ w^-$  with mixing angle $\phi$. 
In that channel, the cross section saturates the S-wave unitarity bound. 
There is no scattering in the orthogonal channel. 
The single-channel-unitarity fixed point is the most natural one for describing a system with a single fine tuning, such as the tuning of the wino mass $M$ to a unitarity value where there is an S-wave resonance at the threshold.

\subsection{Power counting}
\label{sec:powercounting}

In order to define a power counting for the effective field theory associated with the single-channel-unitarity fixed point, we give an explicit parametrization of the transition amplitudes $\mathcal{A}_{ij}(E)$. We introduce two 2-dimensional unit vectors that depend on the mixing angle $\phi$:
\begin{equation}
\label{eq:u,v-def}
\bm{u}(\phi) = \binom{\cos\phi}{\sin\phi}, \qquad \bm{v}(\phi) = \binom{-\sin\phi}{~~\cos\phi}.
\end{equation}
We use these vectors to define two projection matrices and another symmetric matrix:
\begin{subequations}
\begin{eqnarray}
\bm{\mathcal{P}}_u(\phi) &=& \bm{u}(\phi)\, \bm{u}(\phi)^T
\label{eq:Pu}
\\ 
\bm{\mathcal{P}}_v(\phi) &=& \bm{v}(\phi)\, \bm{v}(\phi)^T
\label{eq:Pv}
\\ 
\bm{\mathcal{P}}_m(\phi) &=& \bm{u}(\phi)\, \bm{v}(\phi)^T + \bm{v}(\phi)\, \bm{u}(\phi)^T
\label{eq:Pm}
\end{eqnarray}
\label{eq:Pu,Pv,Pm}%
\end{subequations}
The superscript $T$ indicates transpose. These matricies form a basis for $2 \times 2$ symmetric matrices and are closed under differentiation:
\begin{subequations}
\begin{eqnarray}
\bm{\mathcal{P}}_u'(\phi) &=& \bm{\mathcal{P}}_m(\phi), 
\label{eq:dPu}
\\ 
\bm{\mathcal{P}}_v'(\phi) &=& -\bm{\mathcal{P}}_m(\phi), 
\label{eq:dPv}
\\ 
\bm{\mathcal{P}}_m'(\phi) &=& -2\, \bm{\mathcal{P}}_u(\phi) +2\, \bm{\mathcal{P}}_v(\phi).
\label{eq:dPm}
\end{eqnarray}
\label{eq:dPu,dPv,dPm}%
\end{subequations}

In Ref.~\cite{Lensky:2011he}, Lensky and Birse diagonalized the RG flow near the single-channel-unitarity fixed point, identifying all the scaling perturbations and their scaling dimensions. 
The scaling perturbations to $\bm{M}^{-1/2} \bm{\mathcal{A}}_s^{-1}(E) \bm{M}^{-1/2}$ have the form $(p^2)^i (\Delta^2)^j$, where $p = \sqrt{ME}$,  $\Delta = \sqrt{2 M \delta}$, and $i$ and $j$ are nonnegative integers, multiplied by either $\bm{\mathcal{P}}_u(\phi)$ or $\bm{\mathcal{P}}_v(\phi)$ or $\bm{\mathcal{P}}_m(\phi)$. 
The scaling dimensions are $-1+2i+2j$ in the $\bm{\mathcal{P}}_u$ channel, $1+2i+2j$ in the $\bm{\mathcal{P}}_v$ channel, and $2i+2j$ in the $\bm{\mathcal{P}}_m$ channel. 
The coefficients of the scaling perturbations can be used to provide a complete parametrization of the short distance amplitude matrix in Eq.~\eqref{eq:Ainverse0C}:
\begin{widetext}
\begin{eqnarray}
\label{eq:AZREFT-RG}
\bm{\mathcal{A}}_s^{-1}(E) = \frac{1}{8\pi} \bm{M}^{1/2} 
\left[ \bigg( \sum_{i,j=0}^\infty c^{(u)}_{ij} (p^2)^i (\Delta^2)^j  \bigg) \bm{\mathcal{P}}_u(\phi)
 + \bigg( \sum_{i,j=0}^\infty c^{(v)}_{ij} (p^2)^i (\Delta^2)^j  \bigg) \bm{\mathcal{P}}_v(\phi) \right.
 \nonumber
 \\
\left.
 + \bigg( \sum_{i,j=0}^\infty c^{(m)}_{ij} (p^2)^i (\Delta^2)^j  \bigg) \bm{\mathcal{P}}_m(\phi) 
 + \bm{K}(E) \right]
\bm{M}^{1/2}.~~~
\end{eqnarray}
\end{widetext}
Unitarity constrains the coefficients of the expansions in powers of $p^2$ and $\Delta^2$ to be real. 
At the fixed point, there is a single relevant operator with scaling dimension $-1$. It corresponds to the  parameter $c^{(u)}_{00}$ in the coefficient of $\bm{\mathcal{P}}_u(\phi)$ in Eq.~\eqref{eq:AZREFT-RG}. 
Since the operator is relevant, the parameter $c^{(u)}_{00}$ must be treated nonperturbatively. 
There is a single marginal operator with scaling dimension 0. 
It corresponds to the  parameter $c^{(m)}_{00}$ in the coefficient of $\bm{\mathcal{P}}_m(\phi)$ in Eq.~\eqref{eq:AZREFT-RG}. 
Because of the identity in Eq.~\eqref{eq:dPu}, an infinitesimal change in this parameter can be compensated by an infinitesimal change in the mixing angle $\phi$.
Thus the parameter $c^{(m)}_{00}$ can be absorbed into the mixing angle $\phi$. 
All the other operators are irrelevant operators with scaling dimensions 1 or higher.  
The corresponding parameters can be treated perturbatively. 
The sums in Eq.~\eqref{eq:AZREFT-RG} can be truncated to include only terms with scaling dimensions below some maximum. 
This truncation defines a field theory with a finite number of parameters. 
By increasing the maximum scaling dimension, we obtain a systematically improvable sequence of field theories. 
They define an effective field theory that we refer to  as {\it zero-range effective field theory} (ZREFT).

If we consider winos with a fixed mass splitting $\delta$, the momentum scale associated with transitions between the neutral and charged channels is also fixed.
Thus the coefficients in each of the sums in Eq.~\eqref{eq:AZREFT-RG} with $i=0$ are not distinguishable and can be absorbed into the coefficients of the powers of $p^2$.
We express the resulting parametrization of the short distance amplitude matrix as
\begin{widetext}
\begin{eqnarray}
\label{eq:AinverseZREFT}
\bm{\mathcal{A}}_s^{-1}(E) = \frac{1}{8\pi} \bm{M}^{1/2} 
\Big[ \big(- \gamma_u + \tfrac12 r_u p^2 + \ldots \big) \bm{\mathcal{P}}_u(\phi)
 + \big(-1/a_v + \ldots \big) \bm{\mathcal{P}}_v(\phi)
 \nonumber
 \\
 + \big(\tfrac12 r_m p^2 + \ldots \big) \bm{\mathcal{P}}_m(\phi) + \bm{K}(E) \Big]
\bm{M}^{1/2} \;,
\end{eqnarray}
\end{widetext}
The matrix of transition amplitudes $\bm{\mathcal{A}}$ is obtained by inverting the matrix in Eq.~\eqref{eq:AinverseZREFT} and inserting it into Eq.~\eqref{eq:Amatrix}.
The power counting of ZREFT is enforced by setting the appropriate scattering parameters to zero after the inverse is found.
At leading order (LO) in $Q/\Lambda$, all parameters in the inverse $\bm{\mathcal{A}}_s$ except for $\gamma_u$ are set to zero.
At next to leading order (NLO) in $Q/\Lambda$, the parameter $r_m$ is set to zero, keeping $\gamma_u$, $r_u$, and $a_v$.
In this paper, we present the results and predictions for wino-wino scattering observables calculated in ZREFT at leading order.

At leading order, the result of inverting the matrix in Eq.~\eqref{eq:AinverseZREFT} and enforcing the power counting is
\begin{equation}
\label{eq:AmatrixLOLu}
\bm{\mathcal{A}}_s(E) = 
\frac{8\pi}{L_u(E)}  \, 
\bm{M}^{-1/2} \, \bm{\mathcal{P}}_u(\phi)\,  \bm{M}^{-1/2} \;,
\end{equation}
where $\bm{\mathcal{P}}_u(\phi)$ is the projection matrix defined in Eq.~\eqref{eq:Pu} and $\bm{M}$ is the diagonal matrix in Eq.~\eqref{eq:Mmatrix}. 
The denominator in Eq.~\eqref{eq:AmatrixLOLu} is
\begin{equation}
L_u(E)=-\gamma_u +\sin^2\phi\, K_1(E)\,  - i \cos^2\phi\, \sqrt{ME} \;,
\label{eq:Lu}
\end{equation}
where $K_1(E)$ is the function of the complex energy $E$ defined in Eq.~\eqref{eq:K1-E}.
The operator $\gamma_u$ in Eq.~\eqref{eq:AmatrixLOLu} is a relevant operator.
It must be treated nonperturbatively in order to generate the dynamic length scale $a_0$.
A simple way to do this, is to express $\gamma_u$ in terms of the inverse scattering length $\gamma_0 = 1/a_0$.
The scattering length is defined by the neutral-wino-pair amplitude evaluated at the neutral-wino-pair threshold:
\begin{equation}
\mathcal{A}_{s,00}(E=0) = - 8\pi a_0 /M \;.
\label{eq:A00-a0}
\end{equation}
The left hand side of Eq.~\eqref{eq:A00-a0} depends on $\gamma_u$. Expressing $a_0 = 1/\gamma_0$ in the right hand side of Eq.~\eqref{eq:A00-a0} gives the relation between $\gamma_u$ and $\gamma_0$:
\begin{equation}
\gamma_0 = (1 +  t_\phi^2)\gamma_u - t_\phi^2\, K_1(0) \;.
\label{eq:a0-gammauLO}
\end{equation}
Eq.~\eqref{eq:a0-gammauLO} can be solved for $\gamma_u$ and plugged into Eq.~\eqref{eq:AmatrixLOLu} to give the final result for the leading order short-distance amplitude matrix in ZREFT:
\begin{equation}
\label{eq:AmatrixLOL0}
\bm{\mathcal{A}}_s(E) = 
 \frac{8\pi}{L_0(E)}  \bm{M}^{-1/2}   
\begin{pmatrix}    ~1~     & t_\phi\\  t_\phi & t_\phi^2 \end{pmatrix}  \bm{M}^{-1/2} \;.
\end{equation}
The denominator is
\begin{equation}
L_0(E) =-\gamma_0  + t_\phi^2\, \big[ K_1(E) - K_1(0) \big]\, + \kappa_0(E) \;,
\label{eq:L0-E}
\end{equation}
where $\Delta = \sqrt{2M\delta}$ and $K_1(E)$ defined in Eq.~\eqref{eq:K1-E}. 
The neutral-wino inverse scattering length $\gamma_0$ vanishes at unitarity and can be accurately approximated by the Pad\'e in Eq.~\eqref{eq:gamma0Pade}.
At leading order, the only free parameter is the mixing angle $\phi$ which can be determined by matching with results calculated in NREFT.

\subsection{Matching at leading order}
\label{sec:matchingLO}

The amplitude matrix in Eq.~\eqref{eq:AmatrixLOL0}, inserted into Eq.~\eqref{eq:Amatrix} gives the leading order result for the transition amplitudes between wino-wino pairs:
\begin{equation}
\label{eq:AmatrixLOC}
\bm{\mathcal{A}}(E) = \bm{\mathcal{A}}_C(E) +
 \frac{8\pi}{L_0(E)}  \bm{W}\bm{M}^{-1/2}   
\begin{pmatrix}    ~1~     & t_\phi\\  t_\phi & t_\phi^2 \end{pmatrix}  \bm{M}^{-1/2}\bm{W} \;,
\end{equation}
where $\bm{\mathcal{A}}_C(E)$ is the Coulomb amplitude matrix in Eq.~\eqref{eq:ACmatrix} and $L_0(E)$ is given in Eq.~\eqref{eq:L0-E}.
If the amplitudes are evaluated at a real center-of-mass energy $E$, they define the S-wave T-matrix $\bm{\mathcal{T}}(E)$.
The entry $\mathcal{T}_{00}(E)$ is defined for any center-of-mass energy $E>0$. 
The other elements of $\bm{\mathcal{T}}(E)$ should be interpreted as zero for center-of-mass energies below the charged-wino-pair threshold at $E=2\delta$.

Wino-wino scattering cross sections $\sigma_{i \to j}$ can be calculated analytically in ZREFT at leading order using the expression for the amplitude matrix in Eq.~\eqref{eq:AmatrixLOC}.
The cross sections, averaged over initial spins and summed over final spins are
\begin{subequations}
\begin{eqnarray}
\sigma_{i \to 0}(E) &=&\frac{M^2}{8\pi}
\big| {\cal T}_{i0}(E) \big|^2 \frac{v_0(E)}{v_i(E)},
\label{eq:sig0E-calT}
\\
\sigma_{i\to 1}(E) &=&\frac{M^2}{4\pi}
\big| {\cal T}_{i1}(E) \big|^2 \frac{v_1(E)}{v_i(E)},
\label{eq:sig1E-calT}
\end{eqnarray}
\label{eq:sigE-calT}%
\end{subequations}
where $v_i(E)$ and $v_j(E)$ are the velocities of the incoming and outgoing winos in Eqs.~\eqref{eq:v0,1-E}. 

The T-matrix $\bm{\mathcal{T}}(E)$ is related to the dimensionless T-matrix $\bm{T}(E)$ obtained by solving the Schr\"odinger equation of NREFT through the relation
\begin{equation}
\frac{1}{2M} \, \bm{v}(E)^{-1/2} \, \bm{T}(E)  \, \bm{v}(E)^{-1/2}= 
\frac{1}{8\pi} \, \bm{M}^{1/2} \, \bm{\mathcal{T}}(E) \, \bm{M}^{1/2} \;,
\label{eq:TNR-TZR}
\end{equation}
where $\bm{v} = \mathrm{diag}\left(v_0(E) \,,\, v_1(E)\right)$ is a diagonal matrix of wino velocities in Eqs.~\eqref{eq:v0,1-E}.
For neutral-wino scattering, the relation is
\begin{equation}
\frac{1}{2Mv_0(E)} T_{00}(E) = \frac{M}{8 \pi}\, {\cal T}_{00}(E) \;.
\label{eq:TNR-TZR00}
\end{equation}
Both T-matrix elements $T_{00}(E)$ of NREFT and $\mathcal{T}_{00}(E)$ have low energy effective range expansions.
The range expansion in NREFT is given in Eq.~\eqref{eq:T00NRinv}.
From Eq.~\eqref{eq:AmatrixLOC}, $\mathcal{T}_{00}(E)$ is
\begin{equation}
{\cal T}_{00}(E) = \frac{8\pi/M}{L_0(E)} \;,
\label{eq:T00LO}
\end{equation}
with $L_0(E)$ given in Eq.~\eqref{eq:L0-E}.
The effective range expansion for $\mathcal{T}_{00}(E)$ is
\begin{eqnarray}
\frac{8\pi/M}{{\cal T}_{00}(E)} =
-\gamma_0  - i p + \tfrac12 r_0 \, p^2+ \tfrac18 s_0 \, p^4  + {\cal O}(p^6) \;.
\label{eq:T00LOinv}
\end{eqnarray}
The coefficients in the expansion are found by expanding the left hand side of Eq.~\eqref{eq:T00LOinv} with the expression for $\mathcal{T}_{00}(E)$ in Eq.~\eqref{eq:T00LO} in powers of the relative momentum $p=\sqrt{ME}$.
The effective range $r_0$ and the shape parameter $s_0$ are
\begin{subequations}
\begin{eqnarray}
r_0 &=& 2 t_\phi^2\, K_1'(0)/M \;,
\label{eq:r0LO}
\\
s_0 &=& 4 t_\phi^2\, K_1''(0)/M^2 \;,
\label{eq:s0LO}
\end{eqnarray}
\label{eq:r0s0LO}%
\end{subequations}
where the function $K_1(E)$ is defined in Eq.~\eqref{eq:K1-E} and the primes denote differentiation with respect to the energy $E$.
Explicitly, the effective range is
\begin{equation}
r_0(M) = - \tan^2\phi \frac{\alpha^2M^2}{2 \Delta^3}
\left[ \psi'(i \eta_0) - \frac{1}{2(i \eta_0)^2} - \frac{1}{i \eta_0} \right] \;,
\label{eq:r0match}
\end{equation}
where $i \eta_0 = - \alpha M/(2 \Delta)$ and $\Delta = \sqrt{2 M \delta}$.
We can match onto NREFT by using Eq.~\eqref{eq:r0match} as the matching condition for the mixing angle $\phi$.
In the next section, we compare the mass and energy dependence of scattering observables calculated in NREFT and those found analytically in ZREFT at leading order.

\section{Predictions of ZREFT at leading order}
\label{sec:LOpredictions}

In this section we determine the accuracy of ZREFT at leading order by comparing quantities determined numerically by solving the Schr\"odinger equation using NREFT and calculated analytically using expressions in Sec.~\ref{sec:ZREFTC}.

\begin{figure}[t]
\centering
\includegraphics[width=0.48\linewidth]{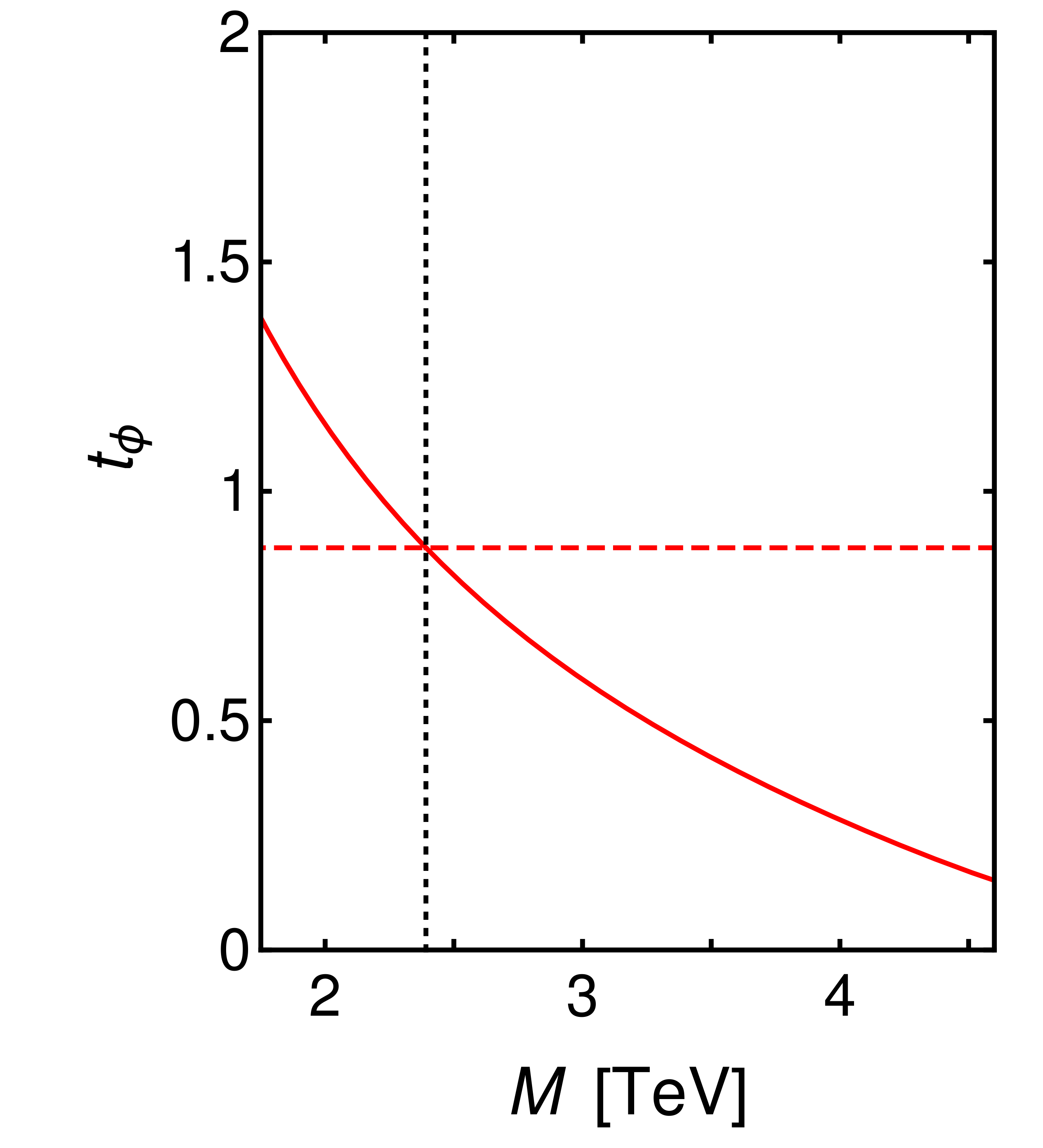}
\includegraphics[width=0.50\linewidth]{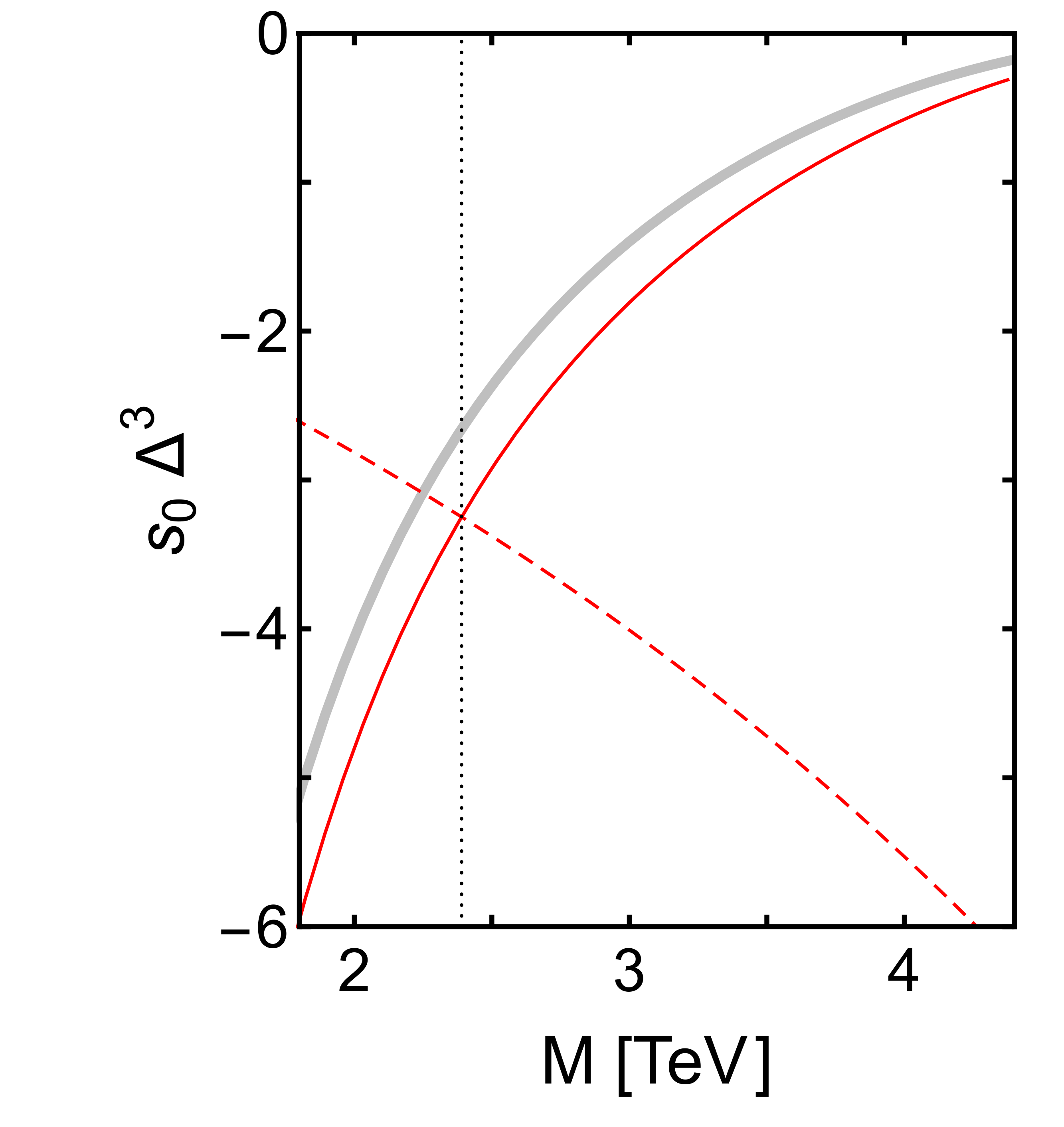}
\caption{Mixing angle $\phi$ (left panel) for ZREFT at LO and neutral-wino shape parameter $s_0$ (right panel) as functions of the wino mass $M$. In the left panel, the parameter $\tan\phi(M)$ (solid red line) is determined from the matching condition for the effective range $r_0$ in Eq.~\eqref{eq:r0match}. The constant value $\tan\phi(M_*)=0.877$ (dashed red line) is determined by matching $r_0$ at unitarity. The shape parameter in the right panel is determined in NREFT (thicker grey line), in ZREFT at LO in Eq.~\eqref{eq:s0LO} with $\tan\phi(M)$ (solid red line) and $\tan \phi(M_*) = 0.877$ (dashed red line). The vertical dotted lines mark the unitarity mass $M_* = 2.39$~TeV.}
\label{fig:tanphis0vsM}
\end{figure}

The result for the leading order S-wave T-matrix in ZREFT is obtained by evaluating Eq.~\eqref{eq:AmatrixLOC} at the center-of-mass energy $E$.
Wino-wino cross sections are calculated in NREFT from Eqs.~\eqref{eq:sigij-T} and in ZREFT from Eqs.~\eqref{eq:sigE-calT}.
The mixing angle is determined by matching the effective range using the result in Eq.~\eqref{eq:r0match}.
At the unitarity mass, the effective range is $r_0(M_*) = -1.653 \sqrt{2M_*\delta}$, giving the mixing angle as $\tan\phi(M_*) = 0.877$.
At other values of the wino mass, the effective range is accurately given by the Pad\'e approximant in Eq.~\eqref{eq:r0Pade}.
The mixing angle is shown as a function of the wino mass in the left panel of Fig.~\ref{fig:tanphis0vsM}.

The neutral-wino shape parameter $s_0$ is shown as a function of the wino mass $M$ in the right panel of Fig.~\ref{fig:tanphis0vsM}.
It is determined in NREFT by solving the Schr\"odinger equation and calculating the effective range expansion in Eq.~\eqref{eq:T00NRinv}.
In ZREFT at leading order, it is found using the expression in Eq.~\eqref{eq:s0LO} along with the value of the mixing angle.
If the matching is carried out as a function of the wino mass, the prediction agrees very well with the NREFT value, as is shown by the solid curve.
If the value of the mixing angle is matching at unitarity and kept constant, the prediction is close at $M_*$, but has an incorrect slope.
Away from the unitarity mass, it is better to determine the mixing angle using the mass-dependent effective range given in Eq.~\eqref{eq:r0Pade}.

\begin{figure}[t]
\centering
\includegraphics[width=0.9\linewidth]{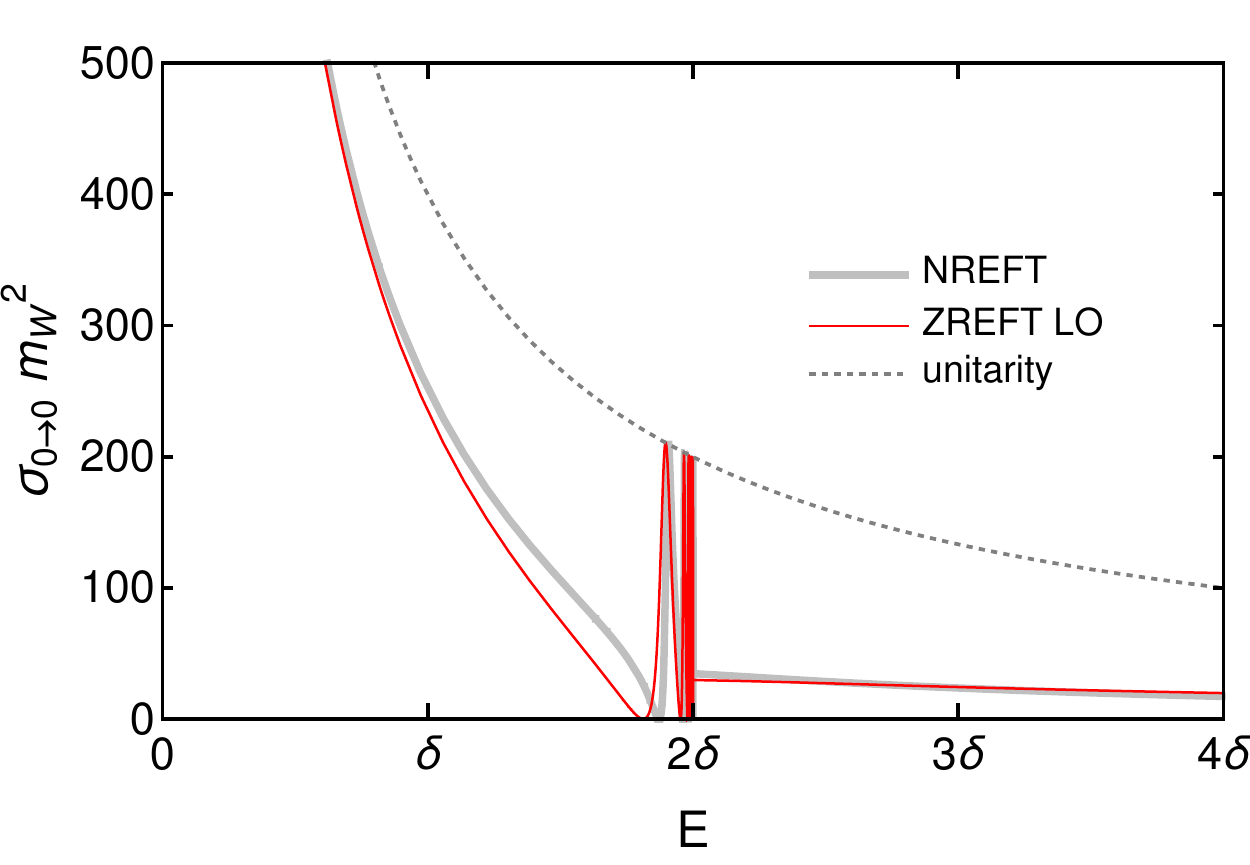}
\caption{Neutral-wino S-wave elastic cross section $\sigma_{0 \to 0}$ as a function of the energy $E$. The cross section at the unitarity mass $M_*=2.39$~TeV is shown for NREFT (thicker grey curve) and for ZREFT at LO with $\tan \phi = 0.877$ (red curve). The S-wave unitarity bound is shown as a dotted curve.}
\label{fig:sigma00vsE-LO}
\end{figure}

The cross sections have the most dramatic energy dependence at a unitarity mass, where $\sigma_{0 \to 0}$ saturates the unitarity bound in Eq.~\eqref{eq:sigma-unitarity0} in the low energy limit.
The cross sections are calculated in NREFT using Eqs.~\eqref{eq:sigij-T}.
The neutral wino S-wave elastic cross section is shown as a function of the center-of-mass energy $E$ in Fig.~\ref{fig:sigma00vsE-LO}.
The leading order ZREFT prediction reproduces all of the important features of the NREFT result, including the resonances just below the charged-wino-pair threshold at $E=2\delta$.

\begin{figure}[t]
\centering
\includegraphics[width=0.48\linewidth]{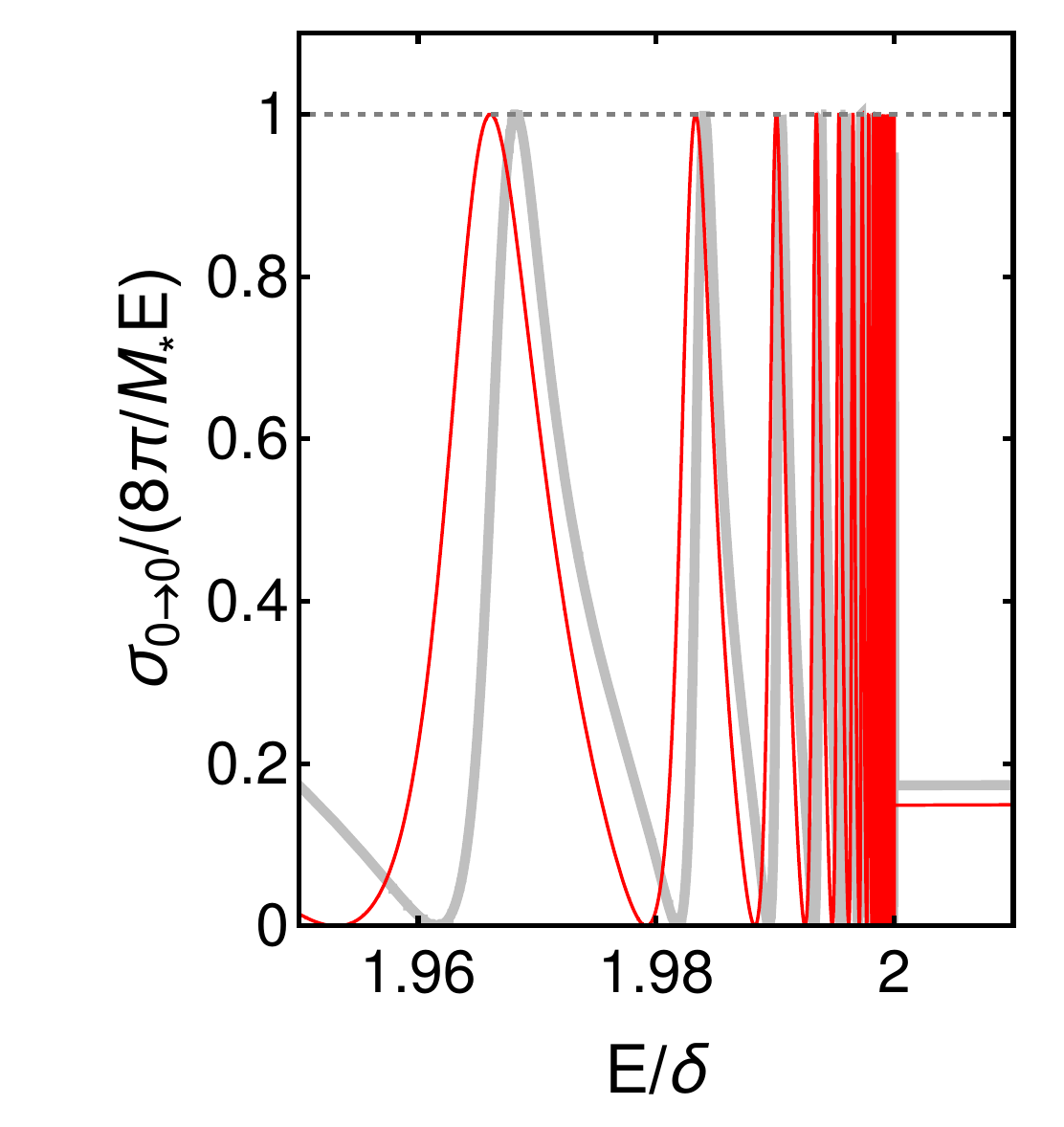}
\includegraphics[width=0.48\linewidth]{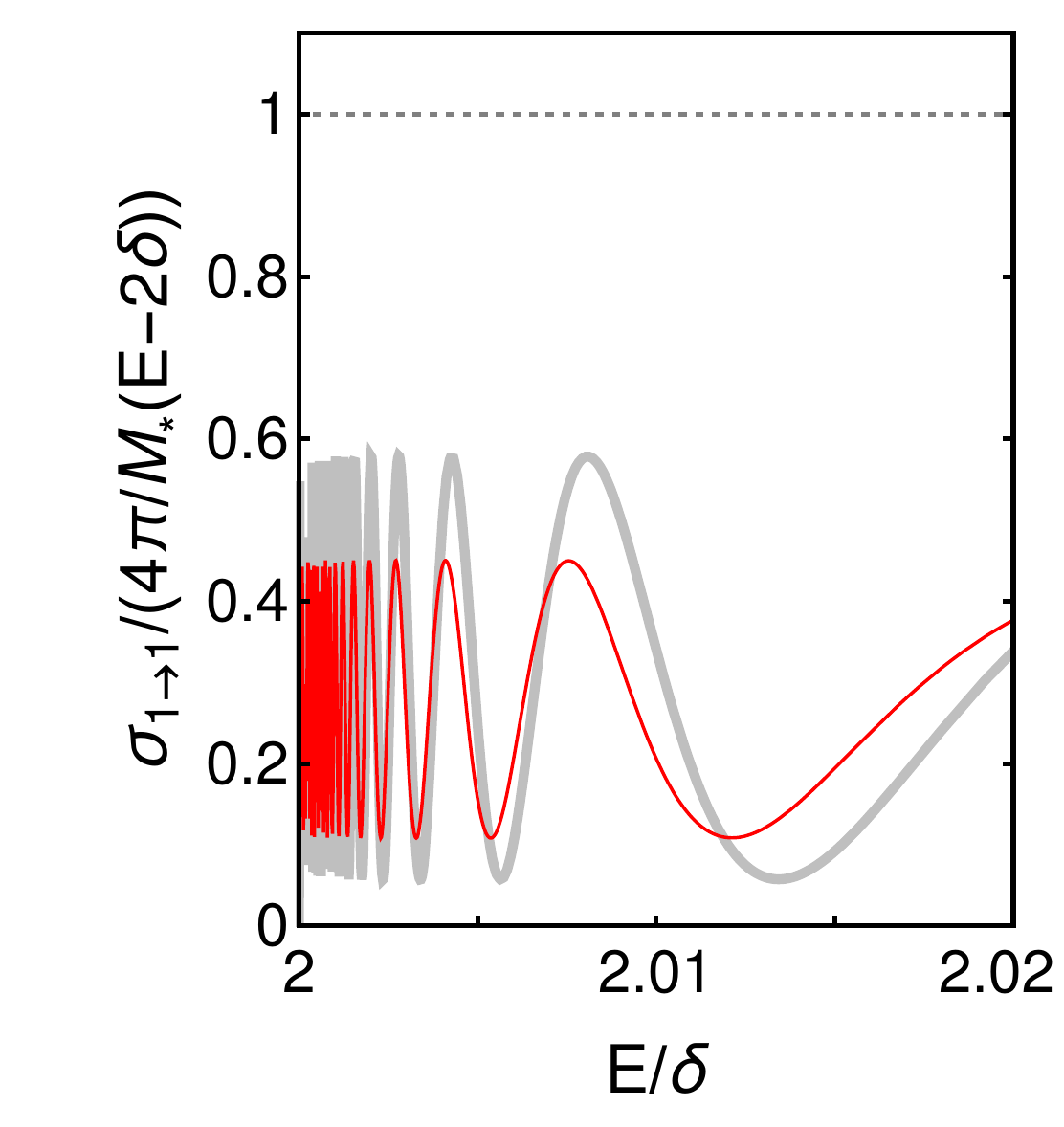}
\caption{Wino-wino S-wave elastic cross sections as a function of the center-of-mass energy $E$ divided by the unitarity bounds in Eq.~\eqref{eq:sigma-unitarity}. The neutral-wino-elastic cross section (left panel) displays resonances at energies below the charged-wino-pair threshold and zeros at $E_n$ in Eq.~\eqref{eq:E-n}. The charged-wino-elastic cross section (right pannel) also has rapid oscilations above threshold which do not saturate the bound. In both panels, the NREFT calculation is shown as a thick gray curve and the ZREFT leading order prediction is shown as a thin red curve.}
\label{fig:elasticwiggles}
\end{figure}
The charged wino S-wave elastic cross section $\sigma_{1 \to 1}$ also displays oscillations in the center-of-mass energy $E$.
These are not resonances since they do not saturate the unitarity bound.
These oscillations are present in the S-wave cross section.
Contributions from other partial waves interfere, and the total cross section has a smooth dependence on the energy.

To examine the resonances in the neutral channel and oscillations in the charged channel, in Fig.~\ref{fig:elasticwiggles}, we divide the elastic cross sections by the respective unitarity bounds in Eq.~\eqref{eq:sigma-unitarity}.
The neutral-wino S-wave elastic cross section in ZREFT displays zeros at the energies $E_n$ in Eq.~\eqref{eq:E-n}.
In NREFT, the locations of the resonances and zeros are slightly increased.
The ZREFT prediction is still an excellent approximation to the cross section.
The charged-wino S-wave elastic cross section does not have resonances or zeros, but has rapid oscillations that increase in frequency as the energy approaches the threshold.
The leading order ZREFT prediction displays the same oscillations, with an amplitude smaller by about 40\%.

\section{Conclusions}
\label{sec:summary}

The Standard Model does not contain a particle that can compose the dark matter.
Models of dark matter are thus models of physics beyond the Standard Model.
One of the most natural ways to extend the Standard Model to include possible dark matter particles is to introduce new multiplets of the Standard Model gauge groups along with some symmetry that forbids their decay into Standard Model particles.
One possibility is to introduce an $SU(2)$ triplet with zero hypercharge so that one component is electrically neutral and can play the role of dark matter.
Such an extension is already included in models of supersymmetry where the new $SU(2)$ triplet is the {\it wino}.
The neutral wino $w^0$ has mass $M$ while the charged winos $w^+$ and $w^-$ have mass $M+\delta$, where $\delta \ll M$.
Winos are charged under the electroweak interactions, which are nonperturbative when $M$ is in the TeV range.
This presents a calculational difficulty in the fundamental quantum field theory describing winos.
Exchanges of weak gauge bosons in wino-wino scattering and annihilation must be summed to all orders in $\alpha_2$.
Photon exchange must be summed to all orders in $\alpha$.

One way around these difficulties is by employing the framework of {\it nonrelativistic effective field theory} (NREFT) in which winos interact through a weak-interaction potential and in which charged winos also interact by exchanging photons.
Cross sections and annihilation rates can be determined by the numerical solution of a Schr\"odinger equation with a potential arrising from the exchange of weak gauge bosons and photons.
At certain values of the wino mass, scattering cross sections and annihilation rates are enhanced by orders of magnitude, as shown in Fig.~\ref{fig:sigma00vsM}.
This ``Sommerfeld enhancement'' occurs because there is a zero-energy resonance at the neutral wino threshold.
Near these resonances, the neutral wino S-wave scattering length $a_0$ is much larger than the range $1/m_W$ of the weak interactions, diverging at the resonance.
The inverse S-wave scattering length $\gamma_0 = 1/a_0$ is shown as a function of the wino mass in Fig.~\ref{fig:gamma0_vsM}.
Exactly at the resonance, the neutral-wino S-wave elastic cross section saturates the unitarity bound in Eq.~\eqref{eq:sigma-unitarity0} as $E \to 0$.
We refer to these masses as {\it unitarity masses}.
This large length scale can be exploited by using an effective field theory in which winos interact through nonperturbative contact interactions and charged winos also interact by exchanging photons.
We refer to this theory as {\it zero-range effective field theory} (ZREFT).
The electromagnetic interaction between charged winos must be treated nonperturbatively by summing up photon exchange to all orders.  

ZREFT is applicable in a range of $M$ where $a_0 > 1/m_W$ and for winos with relative momentum less than $m_W$.
This inequality holds near unitarity masses $M_*$ where $a_0$ diverges.
The first unitarity mass is $M_*=2.39$~TeV.
The range of $M$ where $a_0 > 1/m_W$, and thus ZREFT is applicable, is $1.75-4.6$~TeV.

While NREFT can describe nonrelativistic winos with any mass $M$, ZREFT has distinct advantages.
Two body observables can be calculated analytically in ZREFT and are described very well at leading order in the ZREFT power counting.
The matrix of transition amplitudes for wino-wino scattering is presented at leading order in Eq.~\eqref{eq:AmatrixLOC}.
It has a single free parameter, a mixing angle $\phi$.
The mixing angle is determined by the matching condition in Eq.~\eqref{eq:r0match}.
The left hand side of Eq.~\eqref{eq:r0match} is the neutral-wino effective range $r_0$.
It is shown as a function of the wino mass in Fig.~\ref{fig:r0vsM} along with an accurate Pad\'e approximation which is given in Eq.~\eqref{eq:r0Pade}.
Using the Pad\'e approximation, the mixing angle can be determined over the range in $M$ where ZREFT is applicable.
The dependence of the mixing angle on $M$ is shown in the left panel of Fig.~\ref{fig:tanphis0vsM}.
The prediction for the neutral-wino shape parameter $s_0$ is shown in the right panel of Fig.~\ref{fig:tanphis0vsM}.
Using the mass dependent mixing angle, ZREFT gives a good approximation for $s_0$.

At unitarity masses where the S-wave scattering length diverges, the elastic cross sections have the most dramatic energy dependence.
The neutral-wino S-wave elastic cross section is shown in Fig.~\ref{fig:sigma00vsE-LO}.
The prediction from ZREFT at leading order agrees very well with the result calculated numerically by solving the Schr\"odinger equation.
An enlargement of the energy region near the charged-wino-pair threshold $E=2\delta$ is shown in the left panel of Fig.~\ref{fig:elasticwiggles}.
The ZREFT result reproduces the dramatic resonances in the S-wave elastic cross section.
The charged-wino S-wave elastic cross section is reproduced qualitatively by the ZREFT leading order result, as shown in the right panel of Fig.~\ref{fig:elasticwiggles}.
The prediction displays the same rapid oscillations found from the numerical result, but with an amplitude smaller by about 40\%.

One of the primary motivations for the development of ZREFT for winos was the ``Sommerfeld enhancement'' of the annihilation of a pair of neutral winos into electroweak gauge bosons. 
Wino-pair annihilation not only provides additional wino-wino scattering channels, but also affects other aspects of the few-body physics for low-energy winos. 
The effects of wino-pair annihilation are usually suppressed by $\alpha_2^2 m_W^2/M^2$, which is roughly $10^{-6}$ for $M$ in the TeV region, but they can be dramatic near  a unitarity mass. 
For example, the neutral-wino elastic cross section does not actually diverge at a unitarity mass, but instead has a very narrow peak as a function of $M$ \cite{Blum:2016nrz}. 
The finite maximum cross section comes from unitarization of the wino-pair annihilation, which has not been taken into account in most previous calculations of the Sommerfeld enhancement factor. 
A naive estimate of the maximum cross section is 6 orders of magnitude higher than the cross section above the charged-wino pair threshold. 
The effects of wino-pair annihilation on low-energy winos can be taken into account in ZREFT by analytically continuing real interaction parameters to complex values. 
ZREFT can be used to provide analytic results for low-energy wino-wino cross sections and for inclusive wino-pair annihilation rates, including the effects of the unitarization of wino-pair annihilation. 
The results will be presented in Re.~\cite{BJZ-Annihilation}.



\end{document}